\documentclass[twocolumn,twoside,nofootinbib,prd,aps,tightenlines,10pt]{revtex4-1}

\usepackage{amssymb}
 
\usepackage{graphicx}
\usepackage{latexsym}
\usepackage{mathrsfs}
\usepackage{calrsfs}
\usepackage{amsfonts}
\usepackage{color}

{\count255=\time\divide\count255 by 60 \xdef\hourmin{\number\count255}
  \multiply\count255 by-60\advance\count255 by\time
  \xdef\hourmin{\hourmin:\ifnum\count255<10 0\fi\the\count255}}

\newcommand{\gsim}{\mbox{\raisebox{-1.ex}{$\stackrel
     {\textstyle>}{\textstyle\sim}$}}}
\newcommand{\lsim}{\mbox{\raisebox{-1.ex}{$\stackrel
     {\textstyle<}{\textstyle \sim}$}}}

\begin{document}

\title{Signatures of the Very Early Universe: \\Inflation, Spatial Curvature and Large Scale Anomalies}

\author{Grigor Aslanyan}\email{g.aslanyan@auckland.ac.nz}
\author{Richard Easther}\email{r.easther@auckland.ac.nz}

\affiliation{Department of Physics, University of Auckland,
  Private Bag 92019, Auckland, New Zealand }

\begin{abstract}
\noindent  A short inflationary phase  may not erase all traces of the primordial universe. Associated observables include both spatial curvature and ``anomalies'' in the microwave background or  large scale structure.   The present curvature $\Omega_{K,0}$ reflects the initial curvature, $\Omega_{K,\mathrm{start}}$, and the  angular size of anomalies  depends on $k_\mathrm{start}$, the comoving horizon size at the onset of inflation. We estimate posteriors for  $\Omega_{K,\mathrm{start}}$ and $k_\mathrm{start}$ using current data and  simulations, and show that if either quantity is measured to have a non-zero value, both are likely to be observable. Mappings from $\Omega_{K,\mathrm{start}}$ and $k_\mathrm{start}$ to present-day observables depend strongly on the primordial equation of state;  $\Omega_{K,0}$ spans ten orders of magnitude for a given $\Omega_{K,\mathrm{start}}$ while a simple and general relationship connects  $\Omega_{K,0}$ and  $k_\mathrm{start}$. We show that current bounds on $\Omega_{K,0}$ imply that if $k_\mathrm{start}$ is measurable, the curvature  was already small when inflation began. Finally, since the energy density  changes slowly during inflation, primordial gravitational wave constraints require that  a short inflationary phase is preceded by a nontrivial pre-inflationary phase with critical implications for the expected value of  $\Omega_{K,\mathrm{start}}$.
\end{abstract}

\maketitle

\section{Introduction}
 
Standard slow-roll inflation in combination with $\Lambda$CDM  generically implies that the present-day spatial curvature, $\Omega_{K,0}$, is undetectably small, and current measurements are consistent with a spatially flat universe \cite{Adam:2015rua}.  However, there is long-standing interest in ``Just Enough" inflation (e.g. Refs~\cite{Ratra:1994vw,Bucher:1994gb,Sasaki:1994yt,Bucher:1995ga,Lyth:1995cw,Yamamoto:1995sw,Linde:1995xm,Cicoli:2014bja,Ramirez:2012gt,Ramirez:2011kk}). These investigations have many motivations, including possible large-scale anomalies in the Cosmic Microwave Background (CMB)  \cite{Bennett:2010jb,Ade:2013nlj} as well as a possible future detection of primordial B-modes in the CMB\footnote{The joint analysis of BICEP2, Keck Array, and Planck data has found no evidence for primordial $B$-modes \cite{Ade:2015tva}. However a tensor-to-scalar ratio of $r\sim0.1$  predicted by  simple inflationary models \cite{Ade:2015oja}, is not excluded, and any significant tensor background would increase the tension between simple primordial power spectra and the observed $\langle TT \rangle$ angular power spectrum at low $\ell$ \cite{Abazajian:2014tqa}.}. More generally, investigations of Just Enough Inflation are necessary for a full understanding of inflationary phenomenology.

Currently, the tightest  constraints on $\Omega_{K,0}$ are  obtained from the Planck 2015 dataset together with Baryon Acoustic Oscillations (BAO), Type Ia Supernovae, and Hubble constant measurements \cite{Planck:2015xua}
\begin{equation}\label{current_omega_bounds}
  100\,\Omega_{K,0}=0.08^{+0.40}_{-0.39}\,.
\end{equation}
Bounds on $|\Omega_{K,0}|$ will improve to a level of $10^{-4}$ with future $21$-cm intensity measurements, while $|\Omega_{K,0}| \sim10^{-5}$ is the effective cosmic variance limit \cite{Kleban:2012ph}.

Spatial curvature affects the trajectories of photons propagating from the surface of last scattering toward the observer. Separately, a short inflationary phase may not  erase all  remnants of the pre-inflationary universe, giving rise to  CMB anomalies at large angular scales \cite{Aslanyan:2013jwa} in addition to detectable curvature. If $k_\mathrm{start}$, the comoving horizon size at the onset of inflation,  is close to the present horizon size, the power spectrum may be modified even if the perturbations are strictly Bunch-Davies. 

Accelerated expansion can begin immediately after the Big Bang but the energy density changes slowly during inflation. Given  bounds on the   gravitational wave background inflation must occur at significantly sub-Planckian energies if the number of e-folds is small. Consequently, Just Enough Inflation is  likely to have a non-trivial pre-inflationary phase whose remnants may be visible today, providing clues to the nature of the Big Bang itself. For instance, bounds on $\Omega_{K,0}$ constrain eternal inflation scenarios \cite{Kleban:2012ph,Guth:2012ww} and models in which the Big Bang is a bubble-nucleation event  \cite{Freivogel:2014hca,Bousso:2014jca,Bousso:2013uia,Garriga:1998px,Freivogel:2005vv,Linde:1999wv,Yamauchi:2011qq}. 

The cosmic variance of $\Omega_{K,0}$ and the precision with which it can be measured has been studied in \cite{Waterhouse:2008vb,Vardanyan:2009ft,Knox:2005hx}. A detectable $\Omega_{K,0}$  may be induced by a local inhomogeneity  \cite{Bull:2013fga} and some  mechanisms which account for  possible CMB anomalies suggest that spatial curvature  may be measurable in the next generation of experiments  \cite{Aslanyan:2013jwa,Abazajian:2014tqa}. 

  Pre-inflationary  remnants obviously  reflect the properties of the  pre-inflationary phase. Different scenarios  for the pre-inflationary universe necessarily have different observational consequences but  $k_\mathrm{start}$ is likely to be a key parameter in all of them.  Likewise,  $\Omega_{K,0}$  depends upon the  pre-inflationary curvature $\Omega_{K,\mathrm{start}}$ and the pre- and post-inflationary evolution of the universe. Nucleosynthesis and neutrino freeze-out are described by well-established  analyses  making well-verified  predictions, and inflation is probed via measurements of the perturbation spectrum.  Conversely, the behaviour of the universe during the so-called {\em primordial dark age\/} is largely unknown and the equation of state during both the pre-inflationary and post-inflationary phases is  unconstrained.

Large scale anomalies,  spatial curvature and Just Enough Inflation  all receive significant attention. However,  these  phenomena are not  widely discussed in combination.  In this paper we examine the full range of phenomenology associated with Just Enough Inflation, and   possible correlations between different observables.   There is a simple relationship between $\Omega_{K,0}$ and $k_\mathrm{start}$ and we show that if  $k_\mathrm{start}$ is measurable from astrophysical observations, the curvature was already small at the beginning of inflation.  Moreover, we point out that Just Enough Inflation is typically preceded by a significant non-inflationary phase. In this case, the pre-inflationary universe  is thus likely to have either significant curvature, a long phase of ``fast roll'' inflation, or is itself fine-tuned. This intuition is incorporated into priors describing models of the pre-inflationary universe,  which we use to compute posteriors for $\Omega_{K,0}$ and $k_\mathrm{start}$ from both  data and simulations. 

This paper is organized as follows. In Section \ref{theory_sec} we derive the theoretical dependence of $\Omega_{K,0}$ on the parameters describing the early universe physics, in Section \ref{constraints_sec} we show how the posteriors for $\Omega_{K,0}$ and $k_\mathrm{start}$  constrain these parameters, and we summarize in Section \ref{summary_sec}.

\begin{figure}[t]
\centering
\includegraphics[width=7cm,height=8cm]{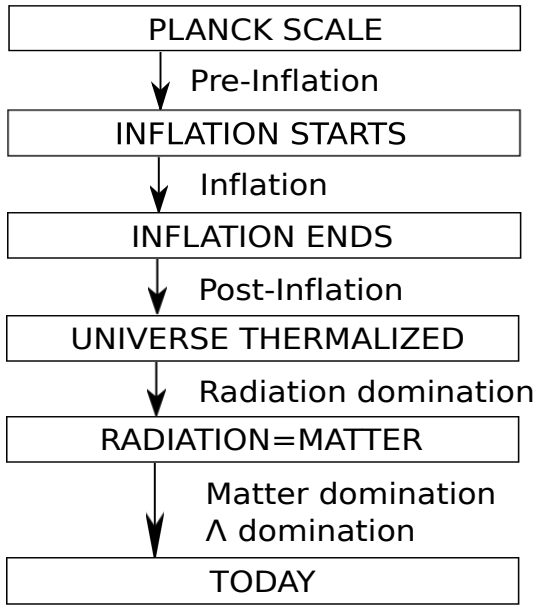}
\vspace{-5pt}
\caption{\label{diagram_fig} The evolution history of the universe.}
\end{figure}

\section{Theoretical Framework}\label{theory_sec}

The overall history of the universe predicted by slow-roll inflation and $\Lambda$CDM is  sketched in Fig. \ref{diagram_fig}. Evolution is assumed to begin at near-Planckian energies. Inflation might  commence immediately after the Big Bang, but the energy density at which cosmologically relevant scales leave the horizon is constrained by bounds on the primordial gravitational wave amplitude. Consequently, Just Enough Inflation must either be preceded by a non-trivial pre-inflationary stage, or the evolution of the universe must begin at a significantly  sub-Planckian energy density. The behaviour of the universe in the period following inflation is likewise unconstrained, but it must  thermalise in order to permit neutrino production and nucleosynthesis. Finally, as is well known, these phases are followed by matter and eventually dark energy domination.

The current scale factor $a_0$ is related to its value at the end of inflation $a_\mathrm{end}$ by the {\it matching equation}\footnote{Throughout this paper the subscript $0$ denotes the current value of a parameter.} \cite{Easther:2011yq}
\begin{eqnarray}\label{matching}
  \ln\left(\frac{a_\mathrm{end}}{a_0}\right)=\frac{1-3w_\mathrm{post}}{12(1+w_\mathrm{post})}\ln\left(\frac{\rho_\mathrm{th}}{\rho_\mathrm{end}}\right)\nonumber\\
  -\frac{1}{4}\ln\left(\frac{\rho_\mathrm{end}}{M_\mathrm{Pl}^4}\right)-71.21 \, .
\end{eqnarray}
Here $\rho_\mathrm{end}$ is the  density at the end of inflation and $M_\mathrm{Pl}$ is the reduced Planck mass.  As in Ref.~\cite{Easther:2011yq} $w_\mathrm{post}$ is the effective equation of state parameter during reheating while the  density $\rho_\mathrm{th}$ is the {\em minimal\/} acceptable thermalization scale. Nucleosynthesis and the cosmic neutrino background put a lower bound on $\rho_\mathrm{th}$ but its specific value forms part of the prior. For a given thermalisation history, the value of $w_\mathrm{post}$ depends on $\rho_\mathrm{th}$ \cite{Easther:2011yq}.      The numerical term in equation~(\ref{matching}) assumes that the effective number of neutrino species is $N_\nu=3.04$. The matching equation ignores the impact of the curvature density on the post-inflationary universe. However,   $|\Omega_{K,0}|\,\lsim\,10^{-2}$,   so even if $\Omega_K$ is not exactly  zero it has no significant effect on the matching between $a_\mathrm{end}$ and $a_0$.

The curvature density is
\begin{equation}\label{curvature_def}
  \Omega_K\equiv-\frac{KM_\mathrm{Pl}^2}{a^2H^2}
\end{equation}
where $K=0,\pm1$  and $H$ is the Hubble parameter. The current value $\Omega_{K,0}$ is related to the value at the end of inflation $\Omega_{K,\mathrm{end}}$ by
\begin{equation}
  \Omega_{K,0}=\Omega_{K,\mathrm{end}}\frac{a_\mathrm{end}^2H_\mathrm{end}^2}{a_0^2H_0^2}\,.
\end{equation}
The curvature  is negligible at the end of inflation, so
\begin{equation}\label{friedmann_end}
  H_\mathrm{end}^2=\frac{\rho_\mathrm{end}}{3M_\mathrm{Pl}^2}\,.
\end{equation}
Putting everything together, we relate $\Omega_{K,0}$ to $\Omega_{K,\mathrm{end}}$ via
\begin{eqnarray}
  \ln\Omega_{K,0}=\ln\Omega_{K,\mathrm{end}}+\frac{1-3w_\mathrm{post}}{6(1+w_\mathrm{post})}\ln\left(\frac{\rho_\mathrm{th}}{\rho_\mathrm{end}}\right)\nonumber\\
  -\frac{1}{2}\ln\left(\frac{\rho_\mathrm{end}}{M_\mathrm{Pl}^4}\right)+\ln\left(\frac{\rho_\mathrm{end}}{3M_\mathrm{Pl}^2H_0^2}\right)-142.42\,.
\end{eqnarray}
Using the standard notation
\[
  H_0=100h\,\frac{\mathrm{km}/\mathrm{s}}{\mathrm{Mpc}}\,,
\]
we get
\begin{eqnarray}\label{curvature_post}
  \ln\Omega_{K,0}=\ln\Omega_{K,\mathrm{end}}+\frac{1-3w_\mathrm{post}}{6(1+w_\mathrm{post})}\ln\left(\frac{\rho_\mathrm{th}}{\rho_\mathrm{end}}\right)\nonumber\\
  +\frac{1}{2}\ln\left(\frac{\rho_\mathrm{end}}{M_\mathrm{Pl}^4}\right)-2\ln h+133.06\,.
\end{eqnarray}

Now focus on the evolution from the Planck scale until the onset of inflation.  Primordial gravitational wave constraints imply that the energy scale of inflation is $2\times 10^{16}$~GeV or less \cite{Ade:2015oja}. We model the pre-inflationary universe as a combination of the inflaton field and a fluid with equation of state parameter $w_\mathrm{pre}$ and  energy density 
\[
  \rho=\rho_\mathrm{pre}\,a^{-3(1+w_\mathrm{pre})}
\]
where $\rho_\mathrm{pre}$ is a constant. The inflaton energy density $\rho_\mathrm{infl}$ is approximately but not exactly constant, as discussed in Section \ref{subsec_infl}. This formalism describes  scenarios with a fast roll phase with $w_\mathrm{pre} \approx -1/3$, as well as conventional, non-inflationary expansion.

Adding all the sources of energy, the pre-inflationary Friedmann equation has the following form
\begin{equation}\label{friedmann_pre}
  H^2=-\frac{KM_\mathrm{Pl}^2}{a^2}+\frac{\rho_\mathrm{infl}}{3M_\mathrm{Pl}^2}+\frac{\rho_\mathrm{pre}}{3M_\mathrm{Pl}^2a^{3(1+w_\mathrm{pre})}}\,.
\end{equation}
Slow roll inflation (as opposed to the generic inflationary criterion of accelerated expansion) starts when the second term on the right hand side of (\ref{friedmann_pre}) becomes  dominant. For concreteness, we define this to occur when $\rho_\mathrm{infl}$ contributes $90\%$ of the energy budget, i.e.
\begin{equation}
\frac{\rho_\mathrm{infl}}{3M_\mathrm{Pl}^2}=0.9H^2\,.
\end{equation}
The  number of $e$-folds following this instant is $N_\mathrm{total}$, 
\begin{equation}
  \ln a_\mathrm{end}=\ln a_\mathrm{start}+N_\mathrm{total}\,.
\end{equation}
In this context, $N_\mathrm{total}$ refers only to the slow roll phase. 

When inflation ends the curvature density $\Omega_{K,\mathrm{end}}$ is related to its initial value $\Omega_{K,\mathrm{start}}$ by 
\begin{equation}
  \Omega_{K,\mathrm{end}}=\Omega_{K,\mathrm{start}}\frac{a_\mathrm{start}^2H_\mathrm{start}^2}{a_\mathrm{end}^2H_\mathrm{end}^2}\,,
\end{equation}
and 
\begin{equation}\label{curvature_infl}
  \ln\Omega_{K,\mathrm{end}}=\ln\Omega_{K,\mathrm{start}}-2N_\mathrm{total}+2\ln\left(\frac{H_\mathrm{start}}{H_\mathrm{end}}\right)\,.
\end{equation}
We  compute $\Omega_{K,\mathrm{start}}$ by solving the pre-inflationary Friedmann equation (\ref{friedmann_pre}) for the scale factor and the Hubble parameter at the onset of inflation;  $\Omega_{K,\mathrm{start}}$ follows via Eq. (\ref{curvature_def}).

The second key quantity is $k_\mathrm{start}$, the largest scale  that crosses the horizon during inflation, or the horizon size at the moment inflation begins. CMB anomalies generated by pre-inflationary relics  depend on the detailed configuration of the pre-inflationary universe, but  $k_\mathrm{start}$ will  determine the current angular size of these anomalies \cite{Aslanyan:2013jwa}. Likewise, if the apparent lack of large scale power in the temperature maps is attributed to a short period of slow-roll inflation \cite{Freivogel:2014hca}, $k_\mathrm{start}$  determines the scale beyond which the  power is suppressed. In comoving coordinates  
\begin{equation}
  k_\mathrm{start,com}=a_\mathrm{start}H_\mathrm{start}\,,
\end{equation}
and its physical value today is
\begin{equation}\label{k_start_def}
  k_\mathrm{start}=\frac{a_\mathrm{start}H_\mathrm{start}}{a_0}=\frac{a_\mathrm{start}}{a_0}\frac{H_\mathrm{start}}{M_\mathrm{Pl}}\cdot3.81\times10^{53}\,\mathrm{Mpc}^{-1}\,.
\end{equation}
We  express $k_\mathrm{start}$ in units of $\mathrm{Mpc}^{-1}$ and with our definitions  
\begin{equation}
  \ln k_\mathrm{start}=-N_\mathrm{total}+\ln\left(\frac{a_\mathrm{end}}{a_0}\right)+\ln\left(\frac{H_\mathrm{start}}{M_\mathrm{Pl}}\right)+123.37\,.
\end{equation}

If the  physical scale corresponding to the horizon size at the onset of inflation is not much larger than the radius of the last scattering surface imprints of pre-inflationary perturbations may be detectable in the CMB  \cite{Aslanyan:2013jwa,Aslanyan:2013zs}. This possibility can be probed via careful analyses of  CMB anomalies. Separately, a  deficit of large scale power in temperature maps may be statistically significant if the CMB $B$ modes prove to have a large  primordial component \cite{Aslanyan:2014mqa,Abazajian:2014tqa}. In this case a new physical scale will become apparent in CMB data and can be mapped to $k_\mathrm{start}$ for specific models of the pre-inflationary epoch.

A simple relationship between $\Omega_{K,0}$, $\Omega_{K,\mathrm{start}}$, and $k_\mathrm{start}$ follows from the definitions (\ref{curvature_def}) and (\ref{k_start_def}), so $\Omega_{K,0}$ is simply connected to $k_\mathrm{start}$ via
\begin{equation}\label{omega_s_k_s}
  \ln\Omega_{K,0}=\ln\Omega_{K,\mathrm{start}}+2\ln k_\mathrm{start}-2\ln h+29.83\,,
\end{equation}
where the last term arises from unit conversions ($k_\mathrm{start}$ is a physical scale expressed in $\mathrm{Mpc}^{-1}$). Critically, this relationship does not depend on the physics of reheating, the inflationary mechanism, its energy scale, or duration. This is a more formal statement of the well-known result that the overall curvature radius of the universe is given by $(aH)^{-1}/\sqrt{|\Omega_K|}$ (see e.g. \cite{Aslanyan:2013jwa}).

\subsection{Priors and Pre-inflationary Evolution}\label{pre_infl_sec}

As noted previously, the main inflationary phase must occur at substantially sub-Planckian energies, given  bounds on the gravitational wave background. Within the Just Enough scenario it is  likely that significant expansion occurs before the onset of slow roll inflation and  $\Omega_{K}$ can evolve during this phase.  

Assuming approximate homogeneity and isotropy, we identify three possible scenarios. The first is that  $w_\mathrm{pre}\approx-1/3$, in which case $\rho\sim 1/a^2$ and $\Omega_{K}$ does not change significantly during the pre-inflationary phase. Conversely, if $w_\mathrm{pre}>-1/3$, the curvature contribution  grows until inflation begins. Given our definitions, $\Omega_{K,\mathrm{start}} \le 0.1$ and if  $w_\mathrm{pre}\gtrsim 0$ this inequality is easily saturated, even if the initial curvature is  small. If both signs of $\Omega_{K}$ are   equally likely and  $\Omega_{K,\mathrm{start}}>0$ and $w_\mathrm{pre}>0$  then we need $\Omega_{K} \ll 1$ initially in order to survive until inflation begins. This  provides an avenue for preferentially producing a universe with negative curvature, independently of the tunnelling argument \cite{Freivogel:2014hca}. Finally, it is possible that the initial state of the universe is effectively set at substantially sub-Planckian scales.

In the light of this analysis, we see three possible priors for $\Omega_{K,\mathrm{start}}$. The first is  $\Omega_{K,\mathrm{start}} = 0.1$, which is the natural choice if the initial curvature can take a large range of values with a substantial phase of non-accelerated expansion before inflation begins; recall that $\Omega_{K}>0$ indicates to a negatively curved universe. Conversely, if it is asserted that the universe undergoes fast-roll inflation with $w_\mathrm{pre}\approx-1/3$ or begins evolving at sub-Planckian densities, we can choose uniform or log priors in which the curvature has either sign, provided $|\Omega_{K,\mathrm{start}}| \le 0.1$.

\begin{figure}[t]
\centering
\includegraphics[width=9cm,height=7cm]{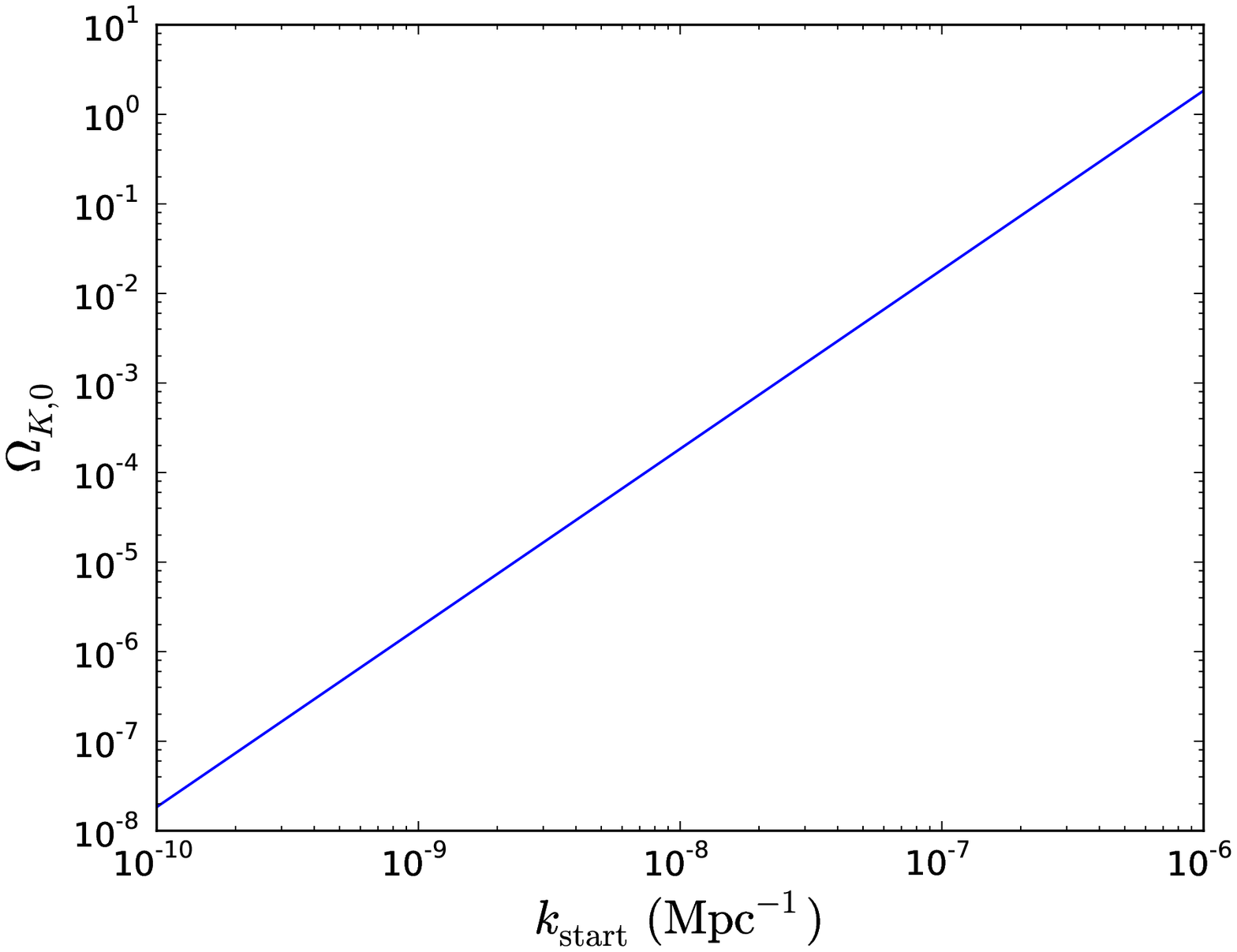}
\vspace{-15pt}
\caption{\label{omega_kstart_fig} The relationship between $\Omega_{K,0}$ and $k_\mathrm{start}$, the largest physical scale that crossed the horizon during inflation; plotted for $\Omega_{K,\mathrm{start}}=0.1$ and $h=0.7$. }
%
\centering
\includegraphics[width=9cm,height=7cm]{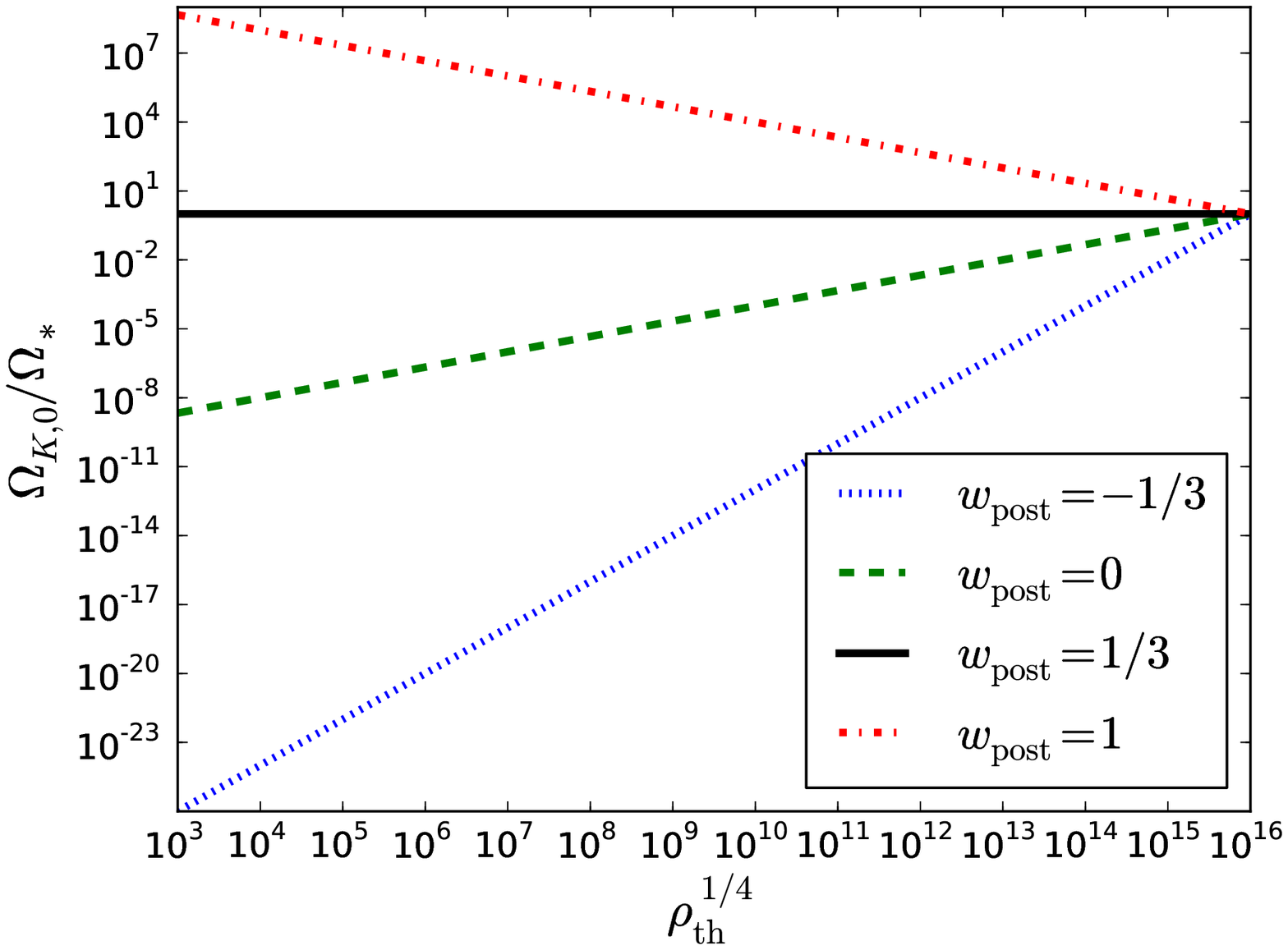}
\vspace{-15pt}
\caption{\label{rho_th_fig} The dependence of $\Omega_{K,0} / \Omega_*$ on the thermalization energy scale $\rho_\mathrm{th}$, where $\Omega_*$ denotes the value of $\Omega_{K,0}$ for the instantaneous entropy generation case.}
\vspace{-15pt}
\end{figure}

Given $k_\mathrm{start}$,  the current spatial curvature  can be  translated into its value at the beginning of inflation mapping constraints on $\Omega_{K,0}$  into constraints on $\Omega_{K,\mathrm{start}}$. We plot the relationship between $\Omega_{K,0}$ on $k_\mathrm{start}$ in Fig. \ref{omega_kstart_fig}, assuming $\Omega_{K,\mathrm{start}}=0.1$ and $h=0.7$.  We see that the current upper bound  $|\Omega_{K,0}|\,\lsim\,10^{-2}$ implies that $k_\mathrm{start}\,\lsim\,10^{-7}\,\mathrm{Mpc}^{-1}$. Consequently,  we can immediately deduce that scenarios in which CMB anomalies are generated by pre-inflationary relics with $k_\mathrm{start}\,\gsim\,10^{-7}\,\mathrm{Mpc}^{-1}$ are in tension with  bounds on $\Omega_{K,0}$, unless $\Omega_{K}$ is already small when inflation begins.

\subsection{Reheating}

Reheating  is not well understood, and many  scenarios have been discussed \cite{Allahverdi:2010xz}. Possibilities include  immediate thermalization, a long matter dominated phase or more exotic scenarios such as cosmic string networks or stiff fluids \cite{Easther:2011yq}.    The Planck   analysis of  inflationary models \cite{Ade:2015oja,Ade:2013uln} lists three representative scenarios which we adopt here:

\begin{enumerate}
  \item {\bf Instantaneous entropy generation}, in which the universe  becomes radiation dominated instantly after inflation.
  \item {\bf Restrictive entropy generation} with $\rho_\mathrm{th}^{1/4}=10^9\,\mathrm{GeV}$, $w_\mathrm{post}\in[-1/3,1/3]$.
  \item {\bf Permissive entropy generation} with $\rho_\mathrm{th}^{1/4}=10^3\,\mathrm{GeV}$, $w_\mathrm{post}\in[-1/3,1]$.
\end{enumerate}
Note that $\rho_\mathrm{th}$ is the scale by which the universe {\em must\/} be  radiation dominated \cite{Easther:2011yq}; actual thermalization can occur at any scale between $\rho_\mathrm{th}$  and  $\rho_\mathrm{end}$.  

We illustrate the dependence of $\Omega_{K,0}$  on the thermalization energy in Fig. \ref{rho_th_fig}. We fix the inflationary energy density at $\rho_\mathrm{end}^{1/4}=10^{16}\,\mathrm{GeV}$ and $\rho_\mathrm{th}^{1/4}$ at $10^3\,\mathrm{GeV}$ and plot $\Omega_{K,0}/ \Omega_{*}$for four different values of $w_\mathrm{post}$: $-1/3$, $0$, $1/3$, and $1$. For the restrictive entropy generation scenario,  $\Omega_{K,0}$ can differ by up to $5$ orders of magnitude  from the instantaneous entropy generation case, while with permissive entropy generation scenario the difference can be up to $24$ orders of magnitude.  For a given inflationary scenario, the required number of e-folds is partially determined by the subsequent expansion history of the universe \cite{Adshead:2010mc} while the curvature is exponentially suppressed during slow-roll inflation. Consequently, a relatively small change in the number of e-folds  leads to a huge variation in  $\Omega_{K,0}$ for fixed $\Omega_{K,\mathrm{start}}$.

\subsection{Inflation}\label{subsec_infl}

If the inflationary scale (and thus $\rho_\mathrm{end}$) is low,  fewer e-folds are needed to suppress the curvature, as is evident from   equation~(\ref{curvature_post}). Thus the answer to the question ``how much inflation is just enough?"  depends on the energy scale at which inflation occurs. 

Using the slow-roll approximation, we  relate the variation of the inflationary energy density to the tensor-to-scalar ratio $r$ and the total number of $e$-folds $N_\mathrm{total}$, correcting the de-Sitter approximation  we used above.  The slow-roll parameter $\epsilon$ is related to the variation of $H$ \cite{Baumann:2009ds}

\begin{equation}
  \frac{d\,\ln H}{d\,N} = -\epsilon\,.
\end{equation}
Assuming that $\epsilon$ is roughly constant during inflation\footnote{This approximation can be improved using the slow-roll hierarchy, and is more accurate when $\epsilon$ is small \cite{Adshead:2008vn}.} we find
\begin{equation}
  \ln\frac{H_\mathrm{start}}{H_\mathrm{end}}=\epsilon N_\mathrm{total}\,.
\end{equation}
Denoting
\begin{equation}
  H_\mathrm{ratio}\equiv\frac{H_\mathrm{start}}{H_\mathrm{end}}\,,
\end{equation}
and using the slow-roll result \cite{Baumann:2009ds} $r \approx 16\epsilon$ gives
\begin{equation}\label{r_N_H_ratio}
  \ln H_\mathrm{ratio}=\frac{r}{16}N_\mathrm{total}\,.
\end{equation}

The last equation will allow us to translate the priors on $r$ and $N_\mathrm{total}$ into a prior for $H_\mathrm{ratio}$. The energy scale of inflation can also be related to $r$ \cite{Baumann:2009ds}. Since we are only interested in slow-roll inflation we will assume that the energy scale at the end of inflation $\rho_\mathrm{end}$ is similar to the pivot scale, for which $r$ is determined. Therefore, we will use \cite{Baumann:2009ds}
\begin{equation}\label{rho_end_r}
  \rho_\mathrm{end}^{1/4}\approx\left(\frac{r}{0.01}\right)^{1/4}\,10^{16}\,\mathrm{GeV}\,.
\end{equation}

\section{Parameter Constraints}\label{constraints_sec}

The current  spatial curvature depends on the detailed history of the evolving universe, making it sensitive to the free parameters that describe the  universe prior to thermalization and radiation domination. Most of these parameters are unknown, so even if $\Omega_{K,0}$ is found to be non-zero, it will not immediately yield constraints on  the detailed properties of the pre-inflationary universe (beyond those that can be learnt from the sign of the curvature) unless other parameters  contributing to the observed value  of $\Omega_{K,0}$ are known  independently.

\begin{table}
\renewcommand{\arraystretch}{1.5}
\setlength{\arraycolsep}{5pt}
\begin{eqnarray*}
\begin{array}{c|c|c}
  \text{Param.} & \text{Prior} & \text{Distribution} \\
\hline
\Omega_{K,\mathrm{start}} & [10^{-10}, 0.1] & \text{Uniform log} \\
w_\mathrm{post} & [-1/3,1/3] & \text{Uniform} \\
\rho_\mathrm{th}^{1/4} & 10^9\,\mathrm{GeV} & \text{Fixed} \\
N_\mathrm{total} & [20,90] & \text{Uniform} \\
h & 0.677\pm0.005 & \text{Gaussian} \\
\end{array}
\end{eqnarray*}
\caption{Parameter priors used in our simulations.}
\label{priors_table}
\end{table}

\begin{table}
\renewcommand{\arraystretch}{1.5}
\setlength{\arraycolsep}{5pt}
\begin{eqnarray*}
\begin{array}{c|cccc}
  \text{Param.} & \text{Sim.}\;1 & \text{Sim.}\;2 & \text{Sim.}\;3 & \text{Sim.}\;4 \\
\hline
100\,\Omega_{K,0} & 0.08\pm0.40 & 0.08\pm0.40 & 0.1\pm0.02 & 0.1\pm0.02 \\
10^6\,k_\mathrm{start} & <10 & 1.0\pm0.2 & <10 & 1.0\pm0.2 \\
r & [10^{-26}, 0.1] & [10^{-26}, 0.1] & [10^{-26}, 0.1] & 0.1\pm0.02 \\
\end{array}
\end{eqnarray*}
\caption{Simulation parameters. Gaussian priors are used for $\Omega_{K,0}$. A uniform prior is used for $k_\mathrm{start}$ for Simulations $1$ and $3$, and a Gaussian prior is used for Simulations $2$ and $4$. A uniform logarithmic prior is used for $r$ for Simulations $1-3$, and a Gaussian prior is used for Simulation $4$.}
\label{params_table}
\end{table}

We now apply the formalism developed here to the problem of extracting early universe parameters from data. We work with both Planck data and simulations for which some or all of $\Omega_{K,0}$, $k_\mathrm{start}$, and $r$ are detectable. 

In this initial treatment the simulations are implemented via a synthetic likelihood function for  $\Omega_{K,0}$ and $k_\mathrm{start}$. In practice, a ``detection" of $k_\mathrm{start}$ would be made with respect to a specific scenario for the pre-inflationary universe; we are simply assuming that such a detection has been made, and  are not performing  analyses based on detailed simulations of the sky.

The priors used in our analysis are summarized in Table \ref{priors_table}. We allow a wide range of values for $\Omega_{K,\mathrm{start}}$ with a uniform logarithmic prior. We use the restrictive entropy generation prior for the reheating parameters. The $N_\mathrm{total}$ prior is consistent with the Planck collaboration analysis of inflationary models \cite{Ade:2015oja,Ade:2013uln}, and the $h$ prior is derived from recent observational bounds \cite{Planck:2015xua}.  For convenience we consider only positive $\Omega_{K,\mathrm{start}}$ (or negative curvature) but the generalisation to a two-sided bound is straightforward. The upper bound on $\Omega_{K,\mathrm{start}}$ follows from our definition. 

We perform  parameter estimations for four scenarios  with different choices for $\Omega_{K,0}$, $k_\mathrm{start}$, and $r$. The first simulation uses the most recent experimental constraints \cite{Planck:2015xua}, while the other simulations assume detections of some of these parameters. Parameter ranges for the different simulations are given in Table \ref{params_table}. We use the Metropolis-Hastings sampler in the numerical library Cosmo++ \cite{Aslanyan:2013opa} for our analysis.

\begin{figure}[t]
\centering
\includegraphics[width=9cm,height=7cm]{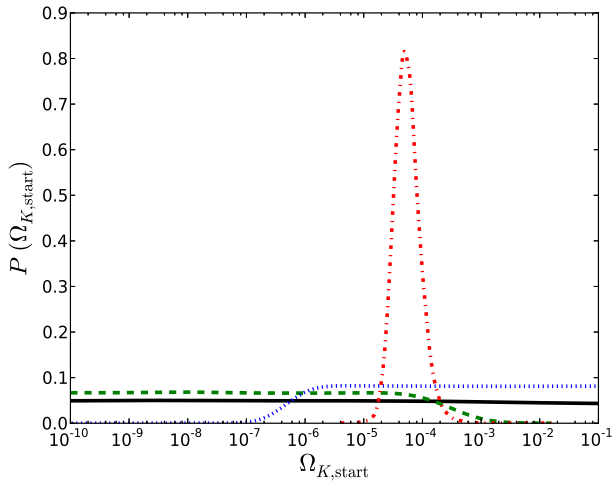}
\vspace{-15pt}
\caption{\label{omega_start_fig} Posteriors of $\Omega_{K,\mathrm{start}}$. The black solid line corresponds to Simulation $1$, the green dashed line to Simulation $2$, and the blue dotted line to Simulation $3$, and the red dash-dotted line to Simulation $4$.}
\vspace{-15pt}
\end{figure}

\begin{figure}[t]
\centering
\includegraphics[width=9cm,height=7cm]{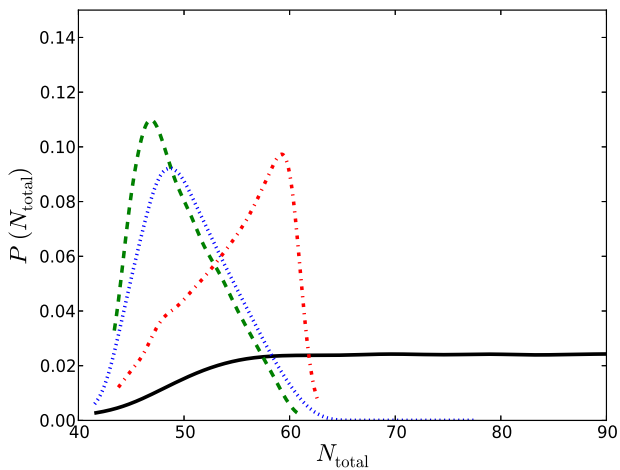}
\vspace{-15pt}
\caption{\label{N_fig} Posteriors of $N_\mathrm{total}$. The labels are the same as in Fig. \ref{omega_start_fig}.}
\vspace{-15pt}
\end{figure}

Simulation 1 is consistent with current experimental bounds and the likelihood incorporates constraints on $\Omega_{K,0}$ estimated from Planck data. Constraints on $k_\mathrm{start}$  depend on a specific model of the pre-inflationary universe and $k_\mathrm{start}$ is only bounded above by present data. For the purposes of this investigation  we take $k_\mathrm{start}<10^{-5}\,\mathrm{Mpc}^{-1}$,  which roughly corresponds to the largest presently observable scales. We assume a uniform logarithmic prior for $r$ with the upper end consistent with current  bounds \cite{Planck:2015xua,Ade:2015tva} and the lower end matching an inflationary energy of $\sim10^{10}\,\mathrm{GeV}$ (see Eq. (\ref{rho_end_r})), consistent with our assumptions about reheating. Simulation 2 is identical to Simulation 1, but assumes a measured $k_\mathrm{start}$ at the $5\,\sigma$ level, with $k_\mathrm{start}=10^{-6}\pm2\times10^{-7}\,\mathrm{Mpc}^{-1}$. Simulation 3, on the other hand, assumes a $5\,\sigma$ measurement of $\Omega_{K,0}$ without a measurement of $k_\mathrm{start}$. Finally, in Simulation 4 we assume that all three parameters $\Omega_{K,0}$, $k_\mathrm{start}$, and $r$ have been detected at the $5\,\sigma$ level.

\begin{figure}[t]
\centering
\includegraphics[width=9cm,height=7cm]{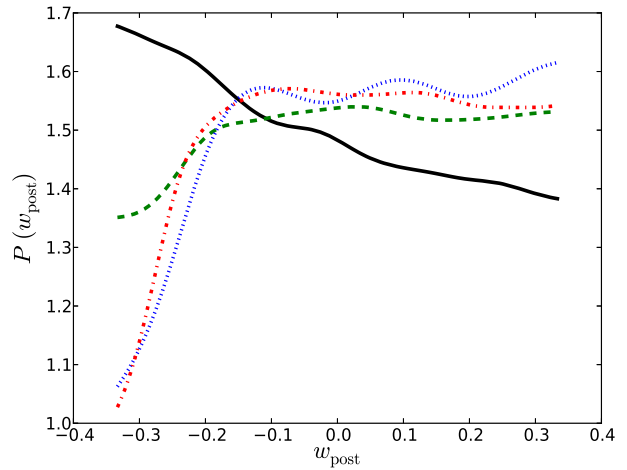}
\vspace{-15pt}
\caption{\label{w_fig} Posteriors of $w_\mathrm{post}$. The labels are the same as in Fig. \ref{omega_start_fig}.}
\vspace{-15pt}
\end{figure}

The  posteriors for $\Omega_{K,\mathrm{start}}$, $N_\mathrm{total}$, and $w_\mathrm{post}$ are shown in Figs \ref{omega_start_fig}, \ref{N_fig}, and \ref{w_fig}, respectively. Current observational bounds (Simulation 1, black solid lines) do not restrict the early universe parameters, yielding a  flat posterior for $\Omega_{K,\mathrm{start}}$ and a lower bound for $N_\mathrm{total}$. The lower bound for $N_\mathrm{total}$ is  consistent with expectations, given that inflation needs to solve the flatness problem which is reflected in $\Omega_{K,0}$ and the horizon problem, reflected in $k_\mathrm{start}$. Unsurprisingly, the posteriors are more restrictive with Simulation 2 (green dashed lines), which assumes a detection of $k_\mathrm{start}$. In this case we get an upper bound on $\Omega_{K,\mathrm{start}}$, and a constraint interval for $N_\mathrm{total}$. The upper bound on $\Omega_{K,\mathrm{start}}$ is easy to understand for this case since we are assuming $\Omega_{K,0}$ has not been detected. On the other hand, if we assume a detection of $\Omega_{K,0}$ instead (Simulation 3, blue dotted lines), we get a lower bound on $\Omega_{K,\mathrm{start}}$, but the posterior of $N_\mathrm{total}$ is similar to the previous case (Simulation 2). Finally, if both $\Omega_{K,0}$ and $k_\mathrm{start}$ are measured (Simulation 4, red dash-dotted line) then the posterior distribution of $\Omega_{K,\mathrm{start}}$ becomes very sharp, as expected from Eq. (\ref{omega_s_k_s}). The posterior distribution of $N_\mathrm{total}$ does not improve significantly, but it shifts slightly as Simulation 4 assumes a measurement of the tensor:scalar ratio, $r$.

Posterior distributions of $w_\mathrm{post}$ are shown in Fig. \ref{w_fig}. Current observational bounds (Simulation 1, black solid line) result in an essentially flat posterior for $w_\mathrm{post}$. A detection of either $k_\mathrm{start}$ (Simulation 2, green dashed line) or $\Omega_{K,0}$ (Simulation 3, blue dotted line) still does not constrain $w_\mathrm{post}$ well. Even if all three parameters $k_\mathrm{start}$, $\Omega_{K,0}$, and $r$ have been detected (Simulation 4, red dash-dotted line), the posterior distribution of $w_\mathrm{post}$ is still too wide to impose significant constraints. A hypothetical detection of $\Omega_{K,0}$ gives a higher preference for larger values of $w_\mathrm{post}$, which is expected since larger values of $w_\mathrm{post}$ imply larger values of $\Omega_{K,0}$ (see Fig. \ref{rho_th_fig}). However, our results show that reheating physics will not be well constrained even if all of $\Omega_{K,0}$, $k_\mathrm{start}$, and $r$ have been measured.\footnote{If the inflationary mechanism is known and the running in the spectral index is not vanishingly small the post-inflationary expansion can be constrained via sensitive measurements of the primordial spectrum \cite{Adshead:2010mc}. }

The analyses presented here have been performed with a single prior.  We have verified that our main conclusions do not change for other reasonable formulations of the prior.  The instantaneous entropy generation scenario yields very similar posteriors for $\Omega_{K,\mathrm{start}}$, so these distributions do not depend on the prior of $w_\mathrm{post}$. The converse is also true  -- a very restrictive prior for $\Omega_{K,\mathrm{start}}$ (for example, assuming that slow roll inflation starts in a curvature dominated universe) does not affect the posterior of $w_\mathrm{post}$. The posterior of $N_\mathrm{total}$, on the other hand becomes narrower if more stringent priors are imposed on either $\Omega_{K,\mathrm{start}}$ or $w_\mathrm{post}$.

\section{Summary}\label{summary_sec}

We have studied how the current curvature density of the universe $\Omega_{K,0}$ is correlated with both the initial curvature and $k_\mathrm{start}$, the largest physical scale  that left the horizon during inflation, and  the evolution of the universe between inflation and thermalisation. The dependence of $\Omega_{K,0}$ on  early universe physics is neatly summarized in Eqs. (\ref{curvature_post}) and (\ref{curvature_infl}) and the relationship between $\Omega_{K,0}$ and $k_\mathrm{start}$ is given in Eq. (\ref{omega_s_k_s}). 


The details of the reheating phase have a  strong impact on $\Omega_{K,0}$. Theoretically plausible reheating scenarios  permit $\Omega_{K,0}$  to span $10$ orders of magnitude for a fixed Just Enough Inflation scenario.  Consequently, while  measuring a non-zero value  of $\Omega_{K,0}$ would have enormous implications for our understanding of the early universe, connecting the measured curvature directly to early universe physics  requires knowledge of the overall expansion history of the universe.

In a scenario with detectable curvature we  expect that the horizon size at the onset of inflation $k_\mathrm{start}$ will be of the same order as the present horizon size. In this case it is plausible that remnants of the pre-inflationary phase will source ``anomalies'' in the CMB or large scale structure \cite{Aslanyan:2013jwa}. If this scale is not much larger than the radius of the last scattering surface then the resulting imprints in the CMB maps may be used to determine $k_\mathrm{start}$.  If both  $k_\mathrm{start}$ and $\Omega_{K,0}$ can be measured then much tighter constraints can be placed on early universe physics. In this context, putative anomalies in the CMB temperature data \cite{Ade:2013nlj} and large scale power suppression \cite{Aslanyan:2014mqa,Abazajian:2014tqa} would provide a route to detecting $k_\mathrm{start}$. The Planck polarization data, in particular, will help to pin down this scale \cite{Aslanyan:2013jwa}. 

Looking at the relationship between $\Omega_{K,0}$ and $k_\mathrm{start}$ we can identify four possible scenarios:
\begin{enumerate}
  \item Both $\Omega_{K,0}$ and $k_\mathrm{start}$ are too small to be detected, in which case there is no evidence for ``Just Enough'' inflation, and our scenario reduces to standard slow-roll inflation.
  \item The initial horizon size $k_\mathrm{start}$ is measured via an analysis of large-angle CMB data and large scale structure information, but $\Omega_{K,0}$ remains consistent with zero.  We see from Eq. (\ref{omega_s_k_s}) that  this scenario requires $\Omega_{K,\mathrm{start}}\ll 1$ (see Fig. \ref{omega_start_fig}),  introducing a pre-inflationary fine-tuning problem. One solution to this dilemma would be a fast-roll phase, which might also provide a plausible mechanism for large scale scalar power suppression \cite{Freivogel:2014hca}.  
  \item A non-zero $\Omega_{K,0}$ is  detected in future experimental data but $k_\mathrm{start}$ is too small to be detected; $\Omega_{K,0}\sim10^{-4}$ will be detectable using the Planck data in combination with future $21$-cm intensity measurements \cite{Kleban:2012ph}. In this case, if the universe is curvature dominated before inflation (implying $\Omega_{K,\mathrm{start}}=0.1$ with our notation), eq. (\ref{omega_s_k_s})  implies $k_\mathrm{start}\,\gsim\,10^{-8}\,\mathrm{Mpc}^{-1}$. This scale is larger than the radius of the last scattering surface by three orders of magnitude.  However, this scenario requires a narrow range of $N_\mathrm{total}$ and $w_\mathrm{post}$ in order to ensure that $\Omega_{K,0}$ is detectable while $k_\mathrm{start}$ corresponds to scales significantly larger than our present horizon.
  \item Both $\Omega_{K,0}$ and $k_\mathrm{start}$ are detected in the  future experimental data. In this case the curvature density of the universe at the onset of inflation, $\Omega_{K,\mathrm{start}}$ can be immediately determined, regardless of the model of reheating (see Eq. (\ref{omega_s_k_s})). However, the physics of reheating will remain poorly constrained without additional information about the inflationary phase itself \cite{Adshead:2010mc}.
\end{enumerate}

Our analysis  shows that either of $\Omega_{K,0}$ and $k_\mathrm{start}$ can be detected without the other, but if  both parameters are measured together substantial information can be gleaned about the early universe. Ideally these parameters would be estimated simultaneously from the data. We have not investigated the correlations between $\Omega_{K,\mathrm{start}}$ and $k_\mathrm{start}$ in specific scenarios of the early universe, but a joint analysis of these two parameters might allow them to be extracted when they are not necessarily detectable with high confidence on their own \cite{Aslanyan:2013jwa}.   Finally, current constraints on $\Omega_{K,0}$ already imply that $k_\mathrm{start}$ corresponds to scales significantly larger than the radius of the surface of last scattering, if $\Omega_{K}\sim1$ when inflation starts. A measurement of  $k_\mathrm{start}$ via large scale anomalies would imply that $\Omega_{K}$ was already small at the onset of slow roll inflation. 

\acknowledgments

We thank Layne Price for  useful discussions. We acknowledge the contribution of the NeSI high-performance computing facilities and the staff at the Centre for eResearch at the University of Auckland. New Zealand's national facilities are provided by the New Zealand eScience Infrastructure (NeSI) and funded jointly by NeSI's collaborator institutions and through the Ministry of Business, Innovation \& Employment's Research Infrastructure programme [{\url{http://www.nesi.org.nz}}].

\bibliography{citations}

\end{document}